\documentclass[a4paper,twocolumn,english,aps,prl,reprint,superscriptaddress,showpacs,showkeys,longbibliography]{revtex4-2}
\usepackage[latin9]{inputenc}
\usepackage{babel}
\usepackage{mathtools}
\usepackage{appendix}
\usepackage{amsmath}
\usepackage{amssymb}
\usepackage{graphicx}
\usepackage{float}
\usepackage{physics}
\usepackage[colorlinks=true, linkcolor=red, citecolor=blue, urlcolor=blue]{hyperref}
\usepackage[usenames,dvipsnames]{color}
\usepackage{bm}
\usepackage{bbold}
\usepackage{braket}
\usepackage{comment}
\usepackage{dcolumn,extarrows}
\usepackage{soul}
\usepackage{footmisc}
\usepackage{pdfpages}
\usepackage{pgf,tikz}
\usepackage{times}

\allowdisplaybreaks[2]

\definecolor{mygreen}{rgb}{0,0.5,0}
\definecolor{myblue}{rgb}{0,0,0.75}
\definecolor{mymagenta}{cmyk}{0,1,0,0.12}

\newcommand{\minus}{
	\setbox0=\hbox{-}
	\vcenter{
		\hrule width\wd0 height \the\fontdimen8\textfont3
	}%
}

\newcommand{\be}{\begin{equation}}\newcommand{\ee}{\end{equation}}
\newcommand{\lf}{\left(}\newcommand{\ri}{\right)}
\newcommand{\Lf}{\left[}\newcommand{\Ri}{\right]}
\newcommand{\la}{\langle}\newcommand{\ra}{\rangle}

\newcommand{\de}{\mathrm{d}}
\newcommand{\hpsi}{\hat{\psi}}

\definecolor{mygreen}{rgb}{0,0.5,0}\definecolor{myblue}{rgb}{0,0,0.75}\definecolor{mymagenta}{cmyk}{0,1,0,0.12}

\newcommand{\ha}{\hat{a}}
\newcommand{\hb}{\hat{b}}
\newcommand{\hH}{\hat{H}}
\newcommand{\bom}{\bar{\omega}}
\newcommand{\mA}{\mathcal{A}}
\newcommand{\mine}{\mathrm{in}}
\newcommand{\macc}{\mathrm{acc}}
\newcommand{\mrot}{\mathrm{rot}}
\newcommand{\mH}{\mathcal{H}}
\newcommand{\mL}{\mathcal{L}}

\makeatletter
\AtBeginDocument{\let\LS@rot\@undefined}
\makeatother

\begin{document}
                  	
\title{Decay of uniformly rotating particles} 

\author{Luciano Petruzziello}\email[]{luciano.petruzziello@uni-ulm.de}
\affiliation{Institut f\"ur Theoretische Physik, Albert-Einstein-Allee 11, Universit\"at Ulm, 89069 Ulm, Germany}
\affiliation{Center for Integrated Quantum Science and Technology (IQST), 89081 Ulm, Germany}
\author{Martin B. Plenio}\email[]{martin.plenio@uni-ulm.de} \affiliation{Institut f\"ur Theoretische Physik, Albert-Einstein-Allee 11, Universit\"at Ulm, 89069 Ulm, Germany}
\affiliation{Center for Integrated Quantum Science and Technology (IQST), 89081 Ulm, Germany}
 
\begin{abstract}
In this paper, we revisit the interpretation of the
circular Unruh effect. To this aim, we rely on the principle of general covariance applied to the decay properties of non-inertial particles. Specifically, we show how the tree-level decay rate of an inverse-$\beta$ process involving scalar fields does not require the introduction of a thermal (or non-thermal) bath in the comoving frame to be a scalar under general coordinate transformations. Instead, we interpret any decay process as an emission of negative-energy quanta, whose existence is motivated by the absence of a global vacuum state for uniformly rotating observers. This implies that, in principle, no uniformly rotating particle can be regarded as stable. 
\end{abstract}
	
\date{\today }
\maketitle
	
\section{Introduction} 

Quantum mechanics and general relativity stand as the main pillars of modern physics: whilst the former explores the behavior and interactions of subatomic particles and molecules in a consistent way, the latter provides a geometric interpretation of gravity and a robust framework to explain the large-scale structure of spacetime. Taken separately, both theories are exceptionally accurate in describing physical manifestations that can be tested with high-precision experiments and observations. However, when trying to reconcile their predictions, no single, universally accepted framework has yet succeeded in unifying them into a coherent scheme. Many candidate models (such as string theory \cite{polchinski,polchinski2}, loop quantum gravity \cite{rovelli,rovelli2} and asymptotic safety \cite{reuter,reuter2} to mention a few) offer promising insights and novel viewpoints, but each of them is affected by conceptual or technical challenges that prevent us from identifying a widely acknowledged quantum description of the gravitational interaction.

On the other hand, it is common belief that, when the quantum fluctuations of the metric tensor embedding the gravitational interaction are negligible, any purported theory of quantum gravity should reduce to quantum field theory in curved spacetime. Despite being a straightforward generalization of the special relativistic theory of quantum fields, this model predicts a highly non-trivial series of new phenomena, whose discovery has had a huge impact on the scientific community. Among these, it is worth recalling black hole evaporation \cite{hawking,hawking2}, cosmological particle creation \cite{cpp,birrell} and the Unruh effect \cite{unruh}.
{As expected}, the experimental conditions required to probe 
{these} effects are extremely demanding. Nevertheless, {considerable effort has been devoted within} the analogue gravity paradigm \cite{analogue}, according to which inaccessible general relativistic regimes can be mimicked in table-top laboratory tests with a suitable preparation of condensed matter and/or classical systems. 

In this regard, the Unruh effect deserves a separate treatment. According to its original formulation, quantum field theory in curved spacetime predicts that uniformly accelerated observers perceive the inertial vacuum as a thermal bath of particles, with a temperature proportional to the proper acceleration \cite{unruhreview}. However, many works also considered its counterpart for {observers in uniform circular motion}
\cite{unruhreview,letaw,davies}, whose non-inertial dynamics {might be expected to} 
give rise to a phenomenon similar to the linearly accelerated version. However, a remarkable difference between the two lies in the fact that the former apparently exhibits a pronounced non-thermal nature, thereby making the identification of the temperature as a function of the acceleration a troublesome task. For this reason, it has been recently suggested \cite{unruhcirc} to revisit the definition of temperature in terms of the excitation and de-excitation rates of the detector, following in the footsteps of Einstein's detailed balance condition \cite{einsteintemp}. Furthermore, another important distinction is that the circular Unruh effect appears to be significantly more suitable for experimental tests, as envisaged in Refs. \cite{bell,bell2,unruhcirc,unruhcirc2,unruhcirc3} and therein.

In this paper, we present a physical argument against the interpretation of the Unruh effect as a non-thermal bath of particles for uniformly circular observers. A first intuition in this direction can be found in Ref. \cite{kl}, where it is argued that the excitation of a rotating detector can occur by emitting negative-energy quanta. The existence of such modes is a consequence of field quantization in the uniformly rotating frame, where the Hamiltonian is not positive-definite and thus unbounded from below. Another way to express this concept consists in acknowledging the impossibility of defining a global vacuum state. At the level of excitations of the field, the non-existence of a global vacuum implies that no particle can really be regarded as stable \cite{unruhreview}.

The last statement suggests an intuitive and efficient path to probe the nature of non-inertial phenomena occurring in a uniformly circular motion. Indeed, the line of reasoning we are going to follow is based upon an interesting result that theoretically validates the presence of the Unruh effect for linearly accelerated observers by consistently relying on the principle of general covariance. Such an idea was firstly hinted at when dealing with the decay properties of accelerated particles \cite{muller}, but it was explicitly implemented only with the analysis of the decay of an accelerated proton \cite{matsas,matsas2,matsas3,suzuki,matsas4,our,our2,our3}. In a nutshell, from the point of view of an inertial observer, a linearly accelerated proton is kinematically allowed to decay {via inverse-$\beta$ emission} 
\be\label{pdec}
p \ \to \ n \ + \ e^+ \ + \ \nu_e\,,
\ee
{because the external field that supplies the acceleration can also bridge the $\sim1.8$ MeV energy gap of this process. The resulting proper lifetime $\tau_p$ is a diffeomorphism scalar, which is why the same value must be obtained in a frame co-accelerating with the proton. In that frame, the proton is at rest, but at the same time immersed in a thermal Unruh bath at temperature $T=\hbar a/2\pi ck_B$. The absorption of quanta from the bath makes the decay possible and reproduces the inertial frame rate. Thus, consistency of quantum field theory demands the Fulling-Davies-Unruh effect.}


By building on the previous idea, one might expect that a similar theoretical test can also be performed in the uniformly circular scenario. A preliminary calculation of the inertial decay rate for the process \eqref{pdec} for rotating protons can be found in \cite{matsascirc}, but there the authors work under specific assumptions and do not explore the possibility of enforcing general covariance to investigate the same process as viewed in the non-inertial frame. In doing this analysis, however, one can prove that there is no need to introduce any ad-hoc thermal (or non-thermal) bath of particles, since the equality between inertial and comoving decay rate can be achieved by allowing the proton to emit negative-energy quanta in the non-inertial frame.

To disclose the above outcome, we organize the paper as follows: In Sec. \ref{rotating} we review the quantization of a massive scalar field for a uniformly rotating observer by closely following \cite{letaw}. Section \ref{proton} explores an inverse-$\beta$ decay-like process involving scalar fields only, as seen in Refs. \cite{muller,our4}, whilst Sec. \ref{twolevel} analyzes the same phenomenon with only a scalar field involved to achieve an analytically closed expression for the decay rate. Finally, Sec. \ref{conclusions}
contains conclusive remarks and outlook.

Henceforth, we are going to use natural units $\hbar=c=1$ and the mostly-negative signature convention for the metric $\mathrm{diag}(+,-,-,-)$.

\section{Unruh effect for rotating observers}\label{rotating}

In order to quantize a neutral massive scalar field for a uniformly rotating observer, it is important to first consider the standard quantization procedure in cylindrical coordinates $(t,r,\theta,z)$. With this choice, the Klein-Gordon equation can be written as \cite{letaw}
\be\label{kg}
\lf\partial_t^2-\partial_r^2-\frac{1}{r}\partial_r-\frac{1}{r^2}\partial_\theta^2-\partial_z^2+M^2\ri\psi=0\,.
\ee
It is possible to show \cite{letaw,davies} that an orthonormal set of modes defined with respect to the Klein-Gordon inner product \cite{birrell} is represented by
\be\label{modes}
u_{q,m,k}=\frac{e^{-i\omega t}e^{i m \theta}e^{i k z}}{2\pi\sqrt{2\omega}}J_m\lf qr\ri,
\ee
with $\omega=\sqrt{q^2+k^2+M^2}$ and $J_\nu(x)$ being the Bessel function of the first kind \cite{grad}. Consequently, a massive scalar field in cylindrical coordinates can be expanded as
\be\label{field}
\hspace{-2.5mm}\hpsi(x)=\sum_{m=-\infty}^\infty\int_0^\infty q \de q\int_{-\infty}^\infty \de k\Lf \ha_{q,m,k}u_{q,m,k}+\ha_{q,m,k}^\dagger u^*_{q,m,k}\Ri\,,
\ee
with $\Lf \ha_{q,m,k},\ha_{q',m',k'}^\dagger\Ri=\delta\lf q-q'\ri\delta\lf k-k'\ri\delta_{m,m'}/q$ and all the other commutators vanishing.

Now, by considering the world line traced by a uniformly rotating motion with angular velocity $\Omega$, it is straightforward to check \cite{letaw,davies} that the Killing vector tangent to the said trajectory is $\chi=\partial_t+\Omega\partial_\theta$, and it is with respect to this quantity that positive and negative frequency solutions must be defined for a comoving observer. In this regard, note that, by writing the components of $\chi$ as $\chi^\mu=(1,0,\Omega,0)$, a straightforward calculation yields $\chi^\mu\chi_\mu=1-r^2\Omega^2$. In principle, there are regions of spacetime where the Killing vector is no longer timelike, which might lead to a non-trivial relation with the annihilation and creation operators of the inertial scenario (where the Killing vector is simply given by $\chi=\partial_t$ \cite{letaw}). To get rid of this ambiguity, one can restrict the quantization in the spatial domain by introducing a boundary condition on the $r$ variable \cite{davies}, in particular requiring that $r<R$, with $R$ such that $R\Omega<1$. Physically, this is achieved by constraining the dynamics of the considered quantum fields inside a cylinder of radius $R$ on which the value of the fields is identically zero. In so doing, $\chi^\mu\chi_\mu>0$ everywhere and the Bogoliubov transformation between inertial and comoving ladder operators is the identity, meaning that both observers agree on the total number of particles. 

On the other hand, if $R$ is such that $R\Omega>1$, the fields span also the region of space beyond the circumference at which the tangential speed is equal to $c$. In this case, the Killing vector is no longer globally timelike since $\chi^\mu\chi_\mu<0$, and thus the definition of positive and negative frequency modes cannot be neatly performed. This means that now also modes with negative frequencies can be part of a physical process, and can be both absorbed and emitted. 

Bearing this in mind, let us quantize a massless scalar field (i.e., $M=0$) according to the new Klein-Gordon equation
\be\label{kg2}
\Lf\lf\partial_t-\Omega\partial_\theta\ri^2-\partial_r^2-\frac{1}{r}\partial_r-\frac{1}{r^2}\partial_\theta^2-\partial_z^2\Ri\psi=0\,,
\ee
which now accounts for the proper Killing vector for a comoving observer. This can be readily achieved by sending $\theta\to\theta-\Omega t\,$ while keeping all the other coordinates unchanged. Because of the Dirichlet boundary condition, the new set of orthonormal modes is given by
\be\label{modes2}
v_{n,m,k,\Omega}=\frac{e^{-i\bom t}e^{i m \theta}e^{i k z}}{2\pi\sqrt{\omega}RJ_{m+1}\lf\xi_{m,n}\ri}J_m\lf\frac{\xi_{m,n} r}{R}\ri,
\ee
where $\bar{\omega}=\omega-m\Omega\,$ is the frequency of the modes with respect to the Killing vector $\chi$, $\xi_{m,n}$ denotes the $n-$th zero of the Bessel function $J_m(x)$ and $\omega=\sqrt{k^2+\xi_{m,n}^2/R^2}$ is the positive frequency one would obtain if the quantization procedure was performed with respect to the globally timelike vector $\partial_t$ (i.e., the Killing vector in the inertial frame).
In light of this, the field in a rotating frame can thus be written as 
\be\label{field2}
\hspace{-2mm}\hpsi(x)=\sum_{{m=-\infty}}^\infty\sum_{n=1}^\infty\int_{-\infty}^\infty \de k\Lf \hb_{n,m,k}v_{n,m,k,\Omega}+\hb_{n,m,k}^\dagger v^*_{n,m,k,\Omega}\Ri\,,
\ee
with $\Lf\hb_{n,m,k},\hb^\dagger_{n',m',k'}\Ri=\delta_{n,n'}\delta_{m,m'}\delta(k-k')$ representing the only non-vanishing commutation relation between creators and annihilators. 

Before concluding this Section, there are two points that need to be addressed. The first point is related to the quantization of a massless scalar field in the inertial frame subjected to the Dirichlet boundary condition $r<R$, which prevents the radius from reaching spatial infinity. Under these conditions, the field expansion resembles the one in Eq. \eqref{field2} with the difference that now the orthonormal modes are given by $u_{n,m,k}=v_{n,m,k,0}$, which formally corresponds to the replacement $\bom\to\omega$.

Finally, we want to emphasize that, when $R\Omega>1$, the quantity $\bom$ can assume negative values. This is a consequence of the fact that $\chi$ is not globally timelike, meaning that there is not a clear distinction between positive and negative frequency states. The non positive-definiteness of $\bom$ will become manifest in the example of the proton decay, which is going to be analyzed in the next Section.

\section{Decay of a non-inertial proton}\label{proton}

Here, we will consider a particle process that mimics the decay of an accelerated proton as seen in Refs. \cite{matsas,matsas2,matsas3,suzuki,matsas4,our,our2,our3} to theoretically support the non-necessity of a thermal or non-thermal bath of particles for uniformly circular observers. For the sake of simplicity, we will work with scalar fields, since the outcome of the analysis does not depend on the nature of the field employed (i.e., boson or fermion), as evident from Ref. \cite{our4}.

Given these premises, let us consider a process analogous to \eqref{pdec} but involving only scalar fields, that is 
\be\label{sdec}
g \ \to \ e \ + \ a_1 \ + \ a_2\,,
\ee
with $g$ and $e$ representing the proton (i.e., ground state) and the neutron (i.e., excited state) of a nucleon system (i.e., two-level system), respectively, whilst $a_1$ and $a_2$ are two particles produced in the decay process which mimic the positron and electron neutrino appearing in \eqref{pdec}. In accordance with \cite{matsas,matsas2,matsas3,suzuki,matsas4,our,our2,our3,our4}, we require the free Hamiltonian of the nucleon to be such that 
\be\label{ham}
\hH|g\ra=E_g|g\ra, \quad \hH|e\ra=E_e|e\ra, \quad E_g<E_e\,.
\ee
In the presence of small centripetal accelerations, a reasonable interacting action for the considered physical setting is given by \cite{matsas}
\be\label{act}
\hat{S}_I=\int \de^4x\sqrt{-g} \ \hat{Q} \ \hpsi_{a_1} \ \hpsi_{a_2},
\ee
with $\hat{Q}$ being the Hermitian monopole
\be\label{Q}
\hat{Q}=e^{i\hH\tau}\hat{q}e^{-i\hH\tau}\delta\lf\vec{x}-\vec{x}(\tau)\ri,
\ee
and $\lambda=|\la g|\hat{q}|e\ra|$ being the effective coupling constant. By assuming the no-recoil approximation, the delta function appearing in \eqref{Q} constrains the motion of the two-level system on a given trajectory $\vec{x}(\tau)$, with $\tau$ being the proper time. For a uniformly circular motion with radius $\rho$ and angular velocity $\Omega$, one has 
\be\label{x}
x^\mu=\lf\gamma\tau,\rho\cos\lf\gamma\Omega\tau\ri,\rho\sin\lf\gamma\Omega\tau\ri,z_0\ri,
\ee
where $\gamma=\lf1-\rho^2\Omega^2\ri^{-1/2}$. Without loss of generality, $z_0$ can be chosen to be zero. With this requirement, the delta function in \eqref{Q} for an inertial observer simply becomes $\delta\lf r-\rho\ri\delta\lf\theta-\Omega t\ri\delta\lf z\ri/r$, while it equals $\delta\lf r-\rho\ri\delta\lf\theta\ri\delta\lf z\ri/r$ in the rotating frame.

Within the aforementioned framework but constraining the motion along the branch of hyperbola representing the trajectory of a linearly accelerated object, one can show that the tree-level proper decay rate $\Gamma_{\mathrm{in}}$ associated with the process \eqref{sdec} is non-vanishing. Since the proper decay rate (i.e., the inverse of the mean proper lifetime) is by construction a scalar under diffeomorphisms, it must be the same for all observers, and in particular also for the one comoving with the two-level system, which sees the nucleon at rest. In this case, the only way for the process to take place is to require the existence of a thermal bath of particles with temperature proportional to the proper acceleration experienced by the two-level system, in accordance with the Unruh effect. In so doing, it is possible to verify \cite{our4} that the decay rate for the comoving observer $\Gamma_{\mathrm{acc}}$ is non-vanishing and it precisely matches the inertial one, namely $\Gamma_{\mathrm{in}}=\Gamma_{\mathrm{acc}}$. Thus, in this context the Unruh effect naturally emerges from the requirement of general covariance of the underlying theory, as it is the only physical occurrence capable of explaining the decay process in the non-inertial reference frame.

Bearing this in mind, should the Unruh effect be similar in nature for the case of uniformly circular motion, it is {reasonable}  
to expect an analogous phenomenon in the frame of the rotating observer. As a first step to check this, let us compute the tree-level transition amplitude $\mA_\mine$ in the inertial frame, which is given by
\be\label{treea}
\mA_\mine=\la e|\otimes\la a_1,a_2|\hat{S}_I|0\ra\otimes|g\ra\,.
\ee
Upon using the expansion \eqref{field2} with the appropriate changes for the inertial frame as discussed in the previous Section since $\rho$ is bounded from above, one then has 
\begin{align}\label{treea2}\nonumber
&\mA_\mine=\lambda\int\de^4x \sqrt{-g} \ e^{i\Delta E \tau} \ \frac{\delta\lf r-\rho\ri\delta\lf\theta-\Omega t\ri\delta\lf z\ri}{r}\\[2mm]
&\hspace{-4mm}\times\sum_{m,m'}\sum_{n,n'}\int^{\infty}_{-\infty}\hspace{-2mm}\de k \ \de k'\la a_1,a_2|\ha^\dagger_{1_{n,m,k}}\ha^\dagger_{2_{n',m',k'}}|0\ra u^*_{n,m,k}u^*_{n',m',k'},
\end{align}
which is the only non-vanishing contribution coming from the product $\hpsi_{a_1}\hpsi_{a_2}$ and where $\Delta E=E_e-E_g>0$. Because of the no-recoil approximation, in full generality we only assume to know the $k$ quantum number of the final state $|a_1,a_2\ra$, meaning that the contraction of the creators appearing in \eqref{treea2} only brings forward the deltas for the momentum along the $z$ axis, meaning $\la a_1,a_2|\ha^\dagger_{1_{n,m,k}}\ha^\dagger_{2_{n',m',k'}}|0\ra=\delta(k-p_1)\delta(k'-p_2)$, while leaving $m,m',n,n'$ completely undetermined.

Next, by opening up the modes $u^*_{n,m,k}u^*_{n',m',k'}$ and recalling that the metric in this case is
\be\label{metric}
\de s^2=\de t^2-\de r^2-r^2\de \theta^2-\de z^2\,,
\ee
from which we deduce $\sqrt{-g}=r$, it is possible to rewrite \eqref{treea2} as
\begin{align}\label{treea3}\nonumber
\mA_\mine&=\frac{\gamma\lambda}{R^2}\sum_{m,m'}\sum_{n,n'}\frac{J_m\lf\frac{\xi_{m,n}\rho}{R}\ri J_{m'}\lf\frac{\xi_{m',n'}\rho}{R}\ri}{J_{m+1}\lf\xi_{m,n}\ri J_{m'+1}\lf\xi_{m',n'}\ri}\\[2mm]
&\times\int_{-\infty}^\infty\frac{\de\tau}{\sqrt{\omega_1\omega_2}} \ e^{i\Lf\Delta E+\gamma\lf\omega_1+\omega_2\ri-\gamma\lf m+m'\ri\Omega\Ri\tau}\,,
\end{align}
with $\omega_1=\sqrt{p_1^2+\xi^2_{m,n}/R^2}$ and $\omega_2=\sqrt{p_2^2+\xi^2_{m',n'}/R^2}$. 

The result of the integral over the proper time is $2\pi\delta\lf\Delta E+\gamma\lf\omega_1+\omega_2\ri-\gamma\lf m+m'\ri\Omega\ri$, meaning that $\mA_\mine$ is an infinite sum over different delta functions. However, depending on the value of the boundary condition $R$, two distinct scenarios can stem from the current analysis. Indeed, if one considers the case in which $R\Omega<1$, it is possible to verify that none of the arguments of the deltas appearing in the sum has a negative value for any choice of $m,m',n,n'$, thus entailing that the total transition amplitude is equal to zero. To show this, consider the quantities $\omega_1$ and $m$, focusing in particular on the combination $\omega_1-m\Omega$ (a similar reasoning holds for $\omega_2$ and $m'$). This difference can never be equal to zero for $m\leq0$, so we should restrict the attention to the case $m>0$. In this case, however, one can resort to the inequality \cite{davies,grad} $\xi_{m,n}>m$ which, together with $R\Omega<1$, can be used to show that $\omega_1-m\Omega>0$. Since in complete analogy one has $\omega_2-m'\Omega>0$ and $\Delta E>0$ by assumption, then it is clear that the argument of the delta in \eqref{treea3} can never be identically zero, thus implying $\mA_\mine=0$.

On the other hand, if $R$ is such that $R\Omega>1$, the above reasoning no longer applies, since there will be values of $m$ for which the argument of the deltas can be negative, thus giving rise to a non-vanishing $\mA_\mine$. This implies that the proton can decay according to the process \eqref{sdec}, and the arguments of the delta functions suggest that the said particle decays into a neutron by emitting two positive-energy quanta of the scalar fields while losing rotational energy. 



At this point, it comes as no surprise that, in the comoving frame, we expect $\mA_\macc\neq0$ by virtue of general covariance. Just like for the linearly accelerated case explored in \cite{matsas,matsas2,matsas3,suzuki,matsas4,our,our2,our3,our4}, there must be some mechanism capable of restoring the equality between the two rates. Since in the rotating frame the proton is at rest, one may think of introducing an ad-hoc bath of particles such that the following three processes are kinematically allowed:
\begin{subequations}
\begin{align}
(i) \ \ \  &g \ + \ a_1 \to \ e \  + \ a_2\,,\\[2mm]
(ii) \ \ \  &g \ + \ a_2 \to \ e \ + \ a_1\,,\\[2mm]
(iii) \ \ \ &g \ + \ a_1 \ + \ a_2 \to \ e\,.
\end{align}
\end{subequations}
In the case of linear acceleration, a proper thermal factor must be introduced next to the probability rate for each of the above processes $\mathcal{P}=n_{\mathrm{BE}}\int\de p_1\de p_2|\mA|^2$ (with $n_{\mathrm{BE}}$ being the Bose-Einstein thermal distribution), motivated by the presence of the Unruh effect. In the uniformly circular case, however, a similar procedure is tricky to implement, as there is no clear indication of what to consider in place of $n_{\mathrm{BE}}$.

For this reason, let us follow a different approach. As a preliminary observation, note that, starting from the metric of the comoving observer
\be\label{metric2}
\de s^2=\lf1-r^2\Omega^2\ri\de t^2+2r^2\Omega\de t\de\theta-\de r^2-r^2\de \theta^2-\de z^2\,,
\ee
and the Lagrangian density for a massless scalar field 
\be\label{lag}
\mL=\frac{\sqrt{-g}}{2}g^{\mu\nu}\partial_\mu\psi\partial_\nu\psi\,,
\ee
one can analyze the shape of the Hamiltonian density $\mH$ defined through the Legendre transformation
\be\label{leg}
\mH=\pi\dot{\psi}-\mL\,, \qquad \pi=\frac{\partial\mL}{\partial\dot{\psi}}\,,
\ee
where the derivative is understood with respect to the Killing vector $\dot{\psi}=\chi\psi$. Note that, when integrated over the whole space, the resulting Hamiltonian yields the same result of the covariant approach employed in \cite{dewitt,gibbons}. A straightforward calculation shows that the Hamiltonian density is equal to
\begin{align}
\nonumber
\mathcal{H} &= \frac{r}{2}\Bigg[
\left(\partial_t\psi\right)^2
 + \left(\partial_r\psi\right)^2
 + \left(\partial_z\psi\right)^2 \\[2mm]
&
 + \frac{1-3 r^2\Omega^2}{r^2}\left(\partial_\theta\psi\right)^2
 + 2\Omega\,\partial_t\psi\,\partial_\theta\psi
 \Bigg]\,.
\label{hamd}
\end{align}
From the above expression, it is evident that the Hamiltonian is no longer bounded from below, meaning that there is no global vacuum state that can be uniquely defined. As a consequence, it is reasonable to assume that, even though the proton $g$ is a stable particle from the point of view of an inertial observer, for a rotating one this is no longer the case. Hence, it is possible to argue that the process \eqref{sdec} also occurs from the standpoint of the comoving observer, despite the fact that $g$ is at rest in this reference system.

By following this line of reasoning, the result of the calculation in the rotating frame is again Eq. \eqref{treea3}, and this holds true because, as already pointed out, we should replace $u_{n,m,k}$ with $v_{n,m,k,\Omega}$, $\omega$ with $\bom$, $\delta\lf\theta-\Omega t\ri$ with $\delta\lf\theta\ri$ and the metric in \eqref{metric} with \eqref{metric2},
for which $\sqrt{-g}=r$. In so doing, one ends up exactly with the same expression, while at the same time avoiding to introduce any ad-hoc thermal bath of particles to let $\Gamma_\mine=\Gamma_\mrot$. 

Differently from the inertial scenario, however, the interpretation concerning the energy balance here has to be revisited. In the currently analyzed case, indeed, the result of the integration over the total proper time is $2\pi\delta\lf\Delta E+\gamma\lf\bom_1+\bom_2\ri\ri$, which is formally equal to the one appearing in \eqref{treea3} because of the implicit definition of $\bom$. Nonetheless, the process \eqref{sdec} here is allowed  not because $g$ loses rotational energy (since it is at rest in the comoving reference frame), but rather because it decays while emitting quanta of the scalar fields with negative energy, which is the only way to let the arguments of the deltas become negative. It is worth stressing once again that this phenomenon arises solely because, in the quantization of the scalar field, the Dirichlet boundary condition imposed on the $r$ variable contemplates the possibility of exceeding the radius at which $r\Omega=1$. Should this not be the case, the decay rate would be equal to zero in both reference frames.

\section{Analytic evaluation of the decay rate}\label{twolevel}

To have a better insight on the behavior of the decay rate as a function of the energy gap between $g$ and $e$, in this Section we will consider a somewhat related but simplified process for which it is possible to analytically compute the expression for $\Gamma$. Without loss of generality, we consider only the calculation in the inertial frame, since in the comoving one the result would be identical.

For this purpose, we now want to compute the transition amplitude $\mA$ for the decay
\be\label{sdec2}
g \ \to \ e \ + \ a\,,
\ee
which in terms of the interacting action is written as $\la e|\otimes\la a|\hat{S}_I|0\ra\otimes|g\ra$. By relying on the same reasoning employed for the process \eqref{sdec}, it is straightforward to see that the transition amplitude $\mA$ equals 
\begin{align}\label{dec}\nonumber
\mA&=\frac{\gamma\lambda}{R}\sum_{m}\sum_{n}\frac{J_m\lf\frac{\xi_{m,n}\rho}{R}\ri }{J_{m+1}\lf\xi_{m,n}\ri}\int_{-\infty}^\infty\frac{\de\tau}{\sqrt{\omega}} \ e^{i\Lf\gamma\lf\omega-m\Omega\ri+\Delta E\Ri\tau}\\[2mm]
&=\frac{2\pi\gamma\lambda}{R}\sum_{m}\sum_{n}\frac{J_m\lf\frac{\xi_{m,n}\rho}{R}\ri }{J_{m+1}\lf\xi_{m,n}\ri}\frac{\delta\lf\gamma\lf\omega-m\Omega\ri+\Delta E\ri}{\sqrt{\omega}}\,,
\end{align}
with $\omega=\sqrt{p^2+\xi^2_{m,n}/R^2}$.

Since we are already summing over all possible configurations related to the quantum numbers $n$ and $m$, the total transition rate is then
\begin{align}\label{dec2}\nonumber
\mathcal{P}&=\int_{-\infty}^\infty\left|\mA\right|^2\de p\\[2mm]\nonumber
&=\frac{4\pi^2\gamma^2\lambda^2}{R^2}\sum_{m,m'}\sum_{n,n'}\frac{J_m\lf\frac{\xi_{m,n}\rho}{R}\ri J_{m'}\lf\frac{\xi_{m',n'}\rho}{R}\ri}{J_{m+1}\lf\xi_{m,n}\ri J_{m'+1}\lf\xi_{m',n'}\ri}\\[2mm]
&\times\int_{-\infty}^\infty\frac{\delta\lf\gamma\lf\omega-m\Omega\ri+\Delta E\ri\delta\lf\gamma\lf\omega'-m'\Omega\ri+\Delta E\ri}{\sqrt{\omega\omega'}}\de p\,.
\end{align}
To evaluate the integral with the two delta functions, we can rewrite each of them using the property
\be\label{delta}
\delta\lf f(x)\ri=\sum_i\frac{\delta\lf x-x_i\ri}{\left|f'\lf x_i\ri\right|}\,,
\ee
where the sum extends over all the roots of $f(x)$ and the prime denotes a derivative with respect to $x$. For the delta functions appearing in \eqref{dec2}, the variable to consider is $p$, and one can check that
\be\label{delta2}
\hspace{-2mm}\delta\lf\gamma\lf\omega-m\Omega\ri+\Delta E\ri=\frac{\left|m\Omega-\frac{\Delta E}{\gamma}\right|\Lf\delta\lf p-p_+\ri+\delta\lf p-p_-\ri\Ri}{\gamma\sqrt{\lf\frac{\Delta E}{\gamma}-m\Omega\ri^2-\frac{\xi_{m,n}^2}{R^2}}},
\ee
with
\be\label{pm}
p_\pm=\pm\sqrt{\lf\frac{\Delta E}{\gamma}-m\Omega\ri^2-\frac{\xi_{m,n}^2}{R^2}}\,.
\ee
Clearly, a similar relation holds for the other delta with the natural replacement $\{m,n\}\to\{m',n'\}$. In light of this, the integral in \eqref{dec2} can be split as the sum of four different integrals, each of which yielding as a result another delta function. However, it is immediate to check that, by collecting these deltas, one has $2 \ \delta\lf p_+-p'_+\ri+2 \ \delta\lf p_++p'_+\ri$, the latter being identically vanishing as the argument can never be equal to zero. On the other hand, the former only gives a contribution when $m=m'$, $n=n'$, but since its value is divergent, it can be rewritten as $2 \ \delta(0)\delta_{m,m'}\delta_{n,n'}$. As argued in \cite{matsas,matsas2,matsas3,suzuki,matsas4,our,our2,our3,matsascirc,our4}, we now identify the quantity $\delta(0)$ as the total proper time
\be\label{pt}
T=\int_{-\infty}^\infty\de\tau=2\pi\delta(0),
\ee
since we are considering an infinite time interval within which the process \eqref{sdec2} takes place. With this knowledge, one can then correctly identify the finite proper decay rate of the particle as $\Gamma=\mathcal{P}/T$, namely
\begin{align}\label{gamma}
\Gamma&=\frac{4\pi\lambda^2}{R^2}\sum_{m}\sum_{n}\frac{J^2_m\lf\frac{\xi_{m,n}\rho}{R}\ri}{J^2_{m+1}\lf\xi_{m,n}\ri}\frac{\left|m\Omega-\frac{\Delta E}{\gamma}\right|}{\lf\frac{\Delta E}{\gamma}-m\Omega\ri^2-\frac{\xi_{m,n}^2}{R^2}}\,,
\end{align}
that is the true invariant quantity under general coordinate transformations.

It is worth observing that, as it should be, the quantity \eqref{gamma} is always positive, since the values of $m$ contributing to the sum are those for which the argument of the delta in \eqref{dec} is negative. These values are such that only the $m$ for which 
\be\label{cond}
m>\frac{\xi_{m,n}}{R\Omega}+\frac{\Delta E}{\Omega\gamma}\,,
\ee
effectively contribute to the sum, and for this range of values the denominator of $\Gamma$ is always positive. 

As a final remark, we stress that the dependence of the decay rate on $\Delta E$ is similar to the response function of an Unruh-Dewitt detector following a uniformly circular trajectory and coupled to a massless scalar field \cite{davies}. Therefore, we expect a seesaw-like behavior for small values of $\Delta E$, in close analogy with the results contained in Refs. \cite{davies,ford}.

\section{Conclusions}\label{conclusions}

In this paper, we have revisited the interpretation of the Unruh effect for uniformly rotating observers. By drawing inspiration from Refs. \cite{davies,kl}, we have explicitly seen how this phenomenon of quantum field theory in curved spacetime unfolds in a realistic (albeit gedanken) scattering process. 

In particular, we have emphasized how, by investigating the decay properties of the inverse-$\beta$ decay-like process \eqref{sdec} involving scalar fields, it is possible to preserve general covariance without resorting to the existence of a thermal bath in the frame where the decaying particle is at rest. This occurrence is in stark contrast with what happens to the uniformly accelerated counterpart, where the Unruh effect is mandatory to achieve the equality between the decay rates in the inertial and the comoving frames \cite{our4}. To investigate the shape of the decay rate a bit further and obtain an analytical result, we have considered the simpler decay process \eqref{sdec2} and showed that, despite the interpretational gap, the prediction is consistent with claims previously appeared in the literature \cite{davies,ford}. Hence, the non-thermal character of the purported circular Unruh effect is rather a signature of the breakdown of the existence of a global vacuum state for comoving observers, which implies the instability of particles traveling in uniform circular orbits.

On a final note, we want to stress that the result supported so far is in line with streamlined considerations that manage to deduce the Unruh effect and temperature from heuristic arguments. Specifically, in Ref. \cite{scard} the Unruh temperature is derived by relating kinetic energy variations with spatial displacements of particles in uniformly accelerated motion. For the uniformly circular trajectory, there is no kinetic energy variation, thereby entailing a vanishing temperature and, more generally, the absence of the effect. A similar result can be achieved also when accounting for the point of view introduced in Ref. \cite{alsing}, where the Unruh effect is regarded as a consequence of a time-dependent Doppler shift between a static observer and a uniformly accelerated particle. For a uniformly circular motion with angular velocity $\Omega$, it is known that the relativistic Doppler effect with respect to a static observer at the center of the trajectory with radius $r$ is proportional to a constant factor $\gamma=\lf1-r^2\Omega^2\ri^{-1/2}$. Hence, one can promptly understand that, because of the constant Doppler shift, the resulting Unruh effect does not exist in this frame.



\bibliography{bib}
\end{document}